\documentclass{article}
\usepackage{ijcai22}

\usepackage{times}

\usepackage{soul}
\usepackage{url}
\usepackage[hidelinks]{hyperref}
\usepackage[utf8]{inputenc}
\usepackage[small]{caption}
\usepackage{graphicx}
\usepackage{amsmath, amsfonts}
\usepackage{microtype}
\usepackage{booktabs}
\usepackage{multirow}
\usepackage{wrapfig}
\title{Towards Robust Unsupervised Disentanglement of Sequential Data --- \\ A Case Study Using Music Audio}

\author{
Yin-Jyun Luo$^1$\footnote{Contact Author}\and
Sebastian Ewert$^2$\And
Simon Dixon$^1$\\
\affiliations
$^1$Centre for Digital Music, Queen Mary University of London\\
$^2$Spotify\\
\emails
yin-jyun.luo@qmul.ac.uk,
sewert@spotify.com,
s.e.dixon@qmul.ac.uk
}

\begin{document}
\newcommand{\todo}[1] {\textcolor{red}{TODO: #1}}
\newcommand{\bv}[1]{\mathbf{#1}}
\newcommand{\pd}[1]{p_\theta(#1)}
\newcommand{\qd}[1]{q_\phi(#1)}
\newcommand{\Dsymbol}[1] {\mathcal{D}_{\mathrm{#1}}}
\newcommand{\KLD}[2] {\Dsymbol{KL}\big(#1 \Vert #2\big)}
\newcommand{\gauss}[2]{\mathcal{N}\big(#1, #2\big)}
\newcommand{\Lo} {\mathcal{L}}
\newcommand{\LB}[1] {\mathcal{L}(\theta, \phi; #1)}
\newcommand{\E}[2] {\mathbb{E}_{#1}\big[ #2 \big]}
\newcommand{\norm}[1] {\Vert #1 \Vert^2}
\graphicspath{ {figures/} }

\newcommand{\fs}[3]{#1^{#2} \rightarrow #1^{#3}}

\newcommand{\refeq}[1] {Eq.~(\ref{#1})}
\newcommand{\figref}[1] {Fig.~\ref{#1}}
\newcommand{\secref}[1] {Section~\ref{#1}}
\newcommand{\tabref}[1] {Table~\ref{#1}}

\maketitle

\begin{abstract}
Disentangled sequential autoencoders (DSAEs) represent a class of probabilistic graphical models that describes an observed sequence with dynamic latent variables and a static latent variable.
The former encode information at a frame rate identical to the observation, while the latter globally governs the entire sequence.
This introduces an inductive bias and facilitates unsupervised disentanglement of the underlying local and global factors.
In this paper, we show that the vanilla DSAE suffers from being sensitive to the choice of model architecture and capacity of the dynamic latent variables, and is prone to collapse the static latent variable.
As a countermeasure, we propose TS-DSAE, a two-stage training framework that first learns sequence-level prior distributions, which are subsequently employed to regularise the model and facilitate auxiliary objectives to promote disentanglement.
The proposed framework is fully unsupervised and robust against the global factor collapse problem across a wide range of model configurations.
It also avoids typical solutions such as adversarial training which usually involves laborious parameter tuning, and domain-specific data augmentation.
We conduct quantitative and qualitative evaluations to demonstrate its robustness in terms of disentanglement on both artificial and real-world music audio datasets.
\end{abstract}

\section{Introduction}
\begin{figure}[!ht]
\centering
 \includegraphics[width=\columnwidth, trim={0 0 0 0}, clip]{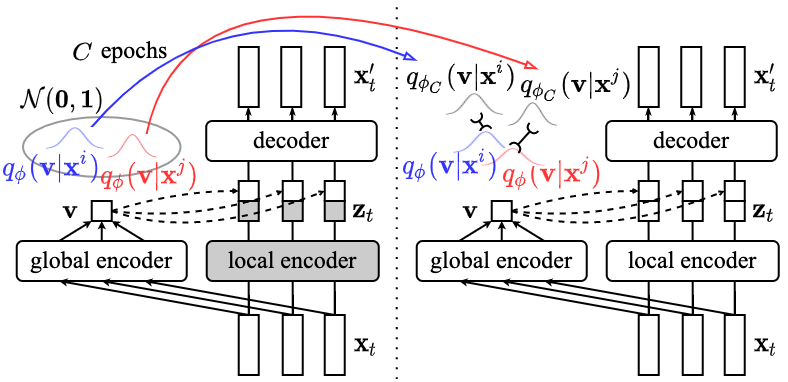}
  \vspace{-5pt}
 \caption{System diagrams of Two-Stage DSAE. Left: The constrained training stage where the local modules are frozen.
 Right: The stage of informed-prior training where the global latent is regularised by the associated posterior learnt from the first stage.
 The dashed arrows denote broadcast along the time-axis.}
 \label{fig:system}
 \vspace{-10pt}
\end{figure}

From a probabilistic point of view, representation learning involves a data generating process governed by multiple explanatory factors of variation~\cite{bengio2013deep}.
The goal of learning a disentangled representation is to extract the underlying factors such that perturbations of one factor only change certain attributes of the observation.
In this sense, disentangled representation promotes model interpretability by exposing semantically meaningful features, and enables controllable data generation by feature manipulation.
 
While supervised learning simplifies training processes, label scarcity for various problems of interest leads to a need for unsupervised techniques. However, as shown by Locatello \textit{et al.}~\shortcite{locatello2018challenging}, disentanglement can only be achieved with either supervision or inductive biases -- and hence any unsupervised system for learning disentangled representations has to involve the latter.
For sequential data, we can aim to disentangle global from local information by leveraging such a structural bias. In this case, the observation is generated by a static (\textit{global}) latent variable associated with the entire sequence, and a series of dynamic (\textit{local}) latent variables varying over time~\cite{hsu2017fhvae,Li2018dsae,Khurana2019fdmm,Zhu2020S3VAESS,Vowels2021VDSMUV,Han2021RWAE,Bai2021cdsvae}.

The disentangled sequential autoencoder (DSAE)~\cite{Li2018dsae} is a minimalistic framework that implements the concept above using a probabilistic graphical model, as illustrated in \figref{fig:dsae}.
However, as we show in \secref{sec:result}, DSAE does not robustly achieve disentanglement but heavily relies on a problem-specific architecture design and parameter tuning.
Several works have built upon DSAE, extending it with either self-supervised learning techniques based on domain-specific data-augmentation~\cite{Bai2021cdsvae}, alternative distance measures for the distributions involved which require extensive hyperparameter tuning or estimations susceptible to the instability resulting from adversarial training~\cite{Han2021RWAE}, or a rather complex parameterisation of a computationally heavy generative model~\cite{Vowels2021VDSMUV}.

In order to improve the robustness of DSAE, we propose TS-DSAE, a simple yet effective framework encompassing a two-stage training method as well as explicit regularisation to improve factor invariance and manifestation. The framework is completely unsupervised and free from any form of data augmentation or adversarial training (but could be combined with either in the future).
We use an artificial as well as a real-world music audio dataset to verify the effectiveness of the proposed framework over a wide range of configurations, and provide both quantitative and qualitative evaluations.
While the baseline models suffer from the collapse of the global latent space, TS-DSAE consistently provides reliable disentanglement (as measured by a classification metric), improves reconstruction quality with increased network capacity without compromising disentanglement, and is able to accommodate multiple global factors shared in the same latent space.


\section{Disentangled Sequential Autoencoders}\label{sec:dsae}

DSAEs~\cite{Li2018dsae,Zhu2020S3VAESS,Bai2021cdsvae,Han2021RWAE,Vowels2021VDSMUV} are a family of probabilistic graphical models representing a joint distribution
\begin{equation}
    \pd{\bv{x}_{1:T}, \bv{z}_{1:T}, \bv{v}} = \\
    p({\bv{v}})
    \prod_{t=1}^{T}\pd{\bv{x}_{t} | \bv{z}_t, \bv{v}}
    \pd{\bv{z}_t | \bv{z}_{<t}},
\end{equation}
where $\bv{x}_{1:T}$ denotes the observed sequence with $T$ time frames, $\bv{z}_{1:T}$ is the sequence of local latent variables, and $\bv{v}$ refers to the global latent variable.
In practice, $\pd{\bv{z}_t | \bv{z_{<t}}} = \gauss{\mu_\theta(\bv{z}_{<t})}{\mathrm{diag}(\sigma_\theta^2(\bv{z}_{<t}))}$ is parameterised by recurrent neural networks (RNNs), and $\pd{\bv{x}_t | \bv{z}_t , \bv{v}}$ is implemented using fully-connected networks (FCNs).
The prior distribution of $\bv{v}$ follows $\gauss{\bv{0}}{\bv{1}}$.
The model is trained to learn separate latent variables $\bv{z}_{1:T}$ and $\bv{v}$ for the local and global factors, respectively, imposing an inductive bias for unsupervised disentanglement, which is otherwise impossible~\cite{locatello2018challenging}.
The uni-modal prior $p(\bv{v})$, however, poses a great challenge to learning an informative latent space, evidenced by our results in \secref{sec:result}.

\begin{figure}[!t]
\centering  
 \includegraphics[width=.8\columnwidth]{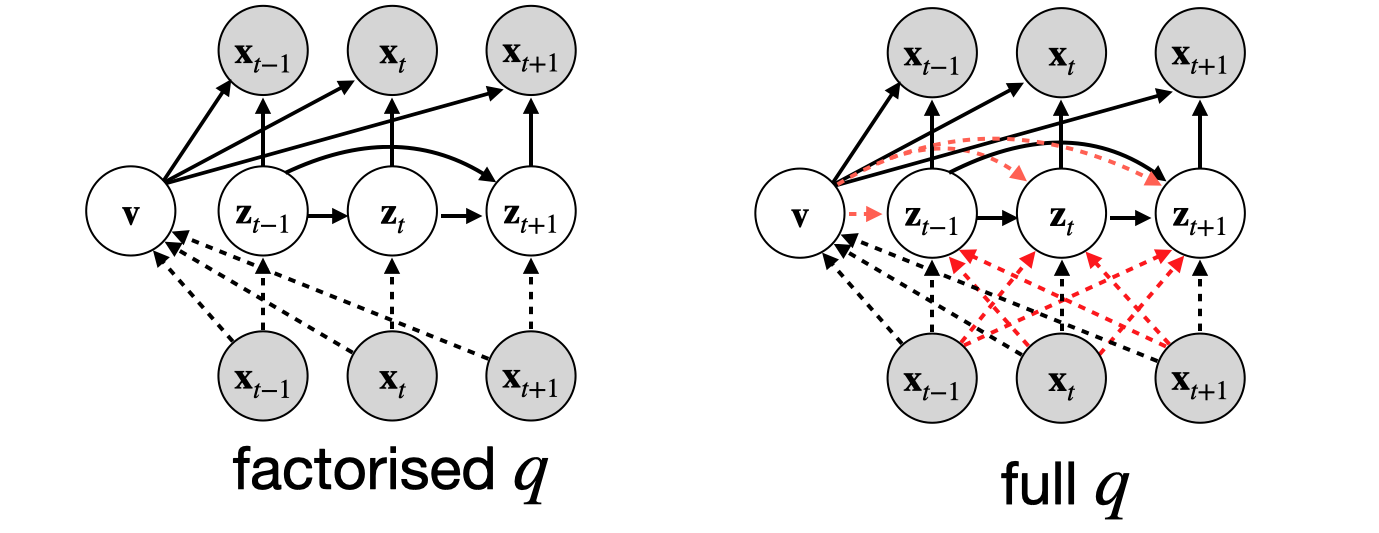}
  \vspace{-5pt}
 \caption{The two models proposed in the original DSAE. The red arrows highlight the enriched inference networks $q_\phi(\cdot)$.}
 \label{fig:dsae}
 \vspace{-10pt}
\end{figure}

Following the framework of variational autoencoders~\cite{kingma2013auto}, inference networks are introduced to optimise the evidence lower bound (ELBO):
\begin{equation}\label{eq:dsae_elbo}
\begin{split}
    &\LB{\bv{x}_{1:T}} \\
    & = \E{\qd{\bv{z}_{1:T}, \bv{v} | \bv{x}_{1:T}}}{
    \log
    \frac{\pd{\bv{x}_{1:T},\bv{z}_{1:T}, \bv{v}}}{\qd{ \bv{z}_{1:T}, \bv{v} | \bv{x}_{1:T}}}
    } \\
    & = \frac{1}{T} \sum_{t=1}^{T}
    \E{\qd{\bv{z}_{t} | \bv{x}_{1:T}, \bv{v}} \qd{\bv{v} | \bv{x}_{1:T}} }{
    \log \pd{\bv{x}_{t} | \bv{z}_{t}, \bv{v}}
    } \\
    & - \frac{1}{T} \sum_{t=1}^{T} 
    \E{\qd{\bv{z}_{<t} | \bv{x}_{1:T}, \bv{v}}}{\KLD{\qd{\bv{z}_t | \bv{x}_{1:T}, \bv{v}}}{\pd{\bv{z}_t | \bv{z}_{<t}}}} \\
    & - \KLD{\qd{\bv{v} | \bv{x}_{1:T}}}{p(\bv{v})}.
\end{split}
\end{equation}
We investigate the two configurations illustrated in \figref{fig:dsae}. 
``full $q$'' follows the inference networks written in \refeq{eq:dsae_elbo}, and $\qd{\bv{z}_t | \bv{x}_{1:T}, \bv{v}}$ can be implemented via RNNs; while ``factorised $q$'' simplifies $\qd{\bv{z}_{1:T} | \bv{x}_{1:T}, \bv{v}} =  \prod_{t=1}^{T}\qd{\bv{z}_t | \bv{x}_t}$ with an FCN shared across the time-axis, which is independent of $\bv{v}$.
In both cases, $\qd{\bv{v}|\bv{x}_{1:T}}$ can be parameterised by either RNNs or FCNs.
We will use ``factorised $q$'' for the exposition in \secref{sec:method}.

A major challenge is that optimising \refeq{eq:dsae_elbo} does not prevent the local latent $\bv{z}_{1:T}$ from capturing all the necessary information for reconstructing the given input sequence $\bv{x}_{1:T}$.
This is referred to as the ``shortcut problem''~\cite{lezama2019jacobian}, where the model completely ignores some latent variables (the global in this case) and only utilises the rest.
In \secref{sec:result}, we show that, without carefully tuning the hyperparameters, the vanilla DSAE is prone to only exploit $\bv{z}_{1:T}$ and ignore $\bv{v}$.

\section{Method}\label{sec:method}
We propose TS-DSAE, which constitutes a two-stage training framework and explicitly imposes regularisation for factor invariance as well as factor rendering in order to encourage disentanglement, as illustrated in \figref{fig:system} which depicts the simplified inference network (factorised $q$) to avoid clutter.

\subsection{Two-Stage Training Framework}\label{sec:two-stage}
The shortcut problem mentioned in \secref{sec:dsae} can be ascribed to the simplicity of the uni-modal prior $p(\bv{v})$ which is not expressive enough to capture the multi-modal global factors, i.e. $\qd{\bv{v}|\bv{x}_{1:T}}$ is over-regularised.
The issue is further exaggerated by the relatively capacity-rich local latent $\bv{z}_{1:T}$ which are allowed to carry information at the frame resolution identical to $\bv{x}_{1:T}$. 
To mitigate the problem, we divide the training into two stages, \textit{constrained training} and \textit{informed-prior training}.

\paragraph{Constrained training.} During constrained training, we freeze some parameters of the local module after initialization including the local encoder and the transition network. This way, the local latents $\bv{z}_t$ resemble random projections from the input and thus are not optimised to hold the most important information to encode the input. That means, we strongly encourage the decoder to focus on the global latent $\bv{v}$ for reconstruction.
As a result, $\qd{\bv{v} | \bv{x}_{1:T}}$ is biased to capture the global factors that are shared across the entire sequence.
From an optimisation perspective, this is equivalent to eliminating the second term (the KL terms for $\bv{z}_t$) from \refeq{eq:dsae_elbo}. 



\paragraph{Informed-prior training.} The training proceeds to the second stage after $C$ epochs of constrained training.
During this stage, all the model parameters are unfrozen and trained regularly using the full objective (\refeq{eq:dsae_elbo}) with a modification.
In particular, instead of setting the global prior to $\gauss{\bv{0}}{\bv{1}}$ as in constrained training, we set:
\begin{equation}
p(\bv{v}^i) = q_{\phi_C}{(\bv{v}^i | \bv{x}^i_{1:T})},
\end{equation}
where $\phi_C$ denotes the parameters of the global encoder at the $C$-th epoch.
That is, we have for \textit{each} input sequence $i$ a corresponding sequence-level prior that has been learnt from constrained training, whereby the last KL term in \refeq{eq:dsae_elbo} is replaced by $\KLD{\qd{\bv{v}^i|\bv{x}^i_{1:T}}}{q_{\phi_C}{(\bv{v}^i | \bv{x}^i_{1:T})}}$.
Note that we differentiate $q_\phi$ from $q_{\phi_C}$ to emphasise that we take a ``snapshot'' of the global encoder $q_{\phi_C}(\cdot)$ at the $C$-th epoch, use the network to parameterise the sequence-specific prior, and continue training the global encoder $q_\phi(\cdot)$ which is initialised by $\phi_C$.
In other words, we keep training the posterior but ``anchor'' the distribution of each sequence $i$ to its associated prior which is the posterior obtained from constrained training and is supposed to capture the sequence-level global factors.

This way, although the local module is introduced over the training, the global latent variables of sequences no longer commonly share the uni-modal prior, thereby mitigating the effect of over-regularisation.

In the next section, we further propose four additional loss terms to encourage disentanglement of the global and local latent variables.

\subsection{Factor Invariance and Manifestation}\label{sec:klds}
Consider the following scheme of inference, replacement, decoding, and inference:
given the \textit{inferred} variables $\bv{z}^i_{1:T} \sim \qd{\bv{z}_{1:T} | \bv{x}^i_{1:T}}$ and $\bv{v}^i \sim \qd{\bv{v} | \bv{x}^i_{1:T}}$, we can \textit{replace} $\bv{v}^i$ with $\bv{v}^j$ inferred from another sequence $j$, and \textit{decode} $\bv{x}_{1:T}^{\bv{v}^i \rightarrow \bv{v}^j} \sim  \pd{\bv{x}_{1:T} | \bv{z}^i_{1:T}, \bv{v}^{j}}$.
We can then \textit{infer} $\bv{z}_{1:T}^{\bv{v}^i \rightarrow \bv{v}^j} \sim \qd{\bv{z}_{1:T} | \bv{x}_{1:T}^{\bv{v}^i \rightarrow \bv{v}^j}}$ and $\bv{v}^{\bv{v}^i \rightarrow \bv{v}^j} \sim \qd{\bv{v} | \bv{x}_{1:T}^{\bv{v}^i \rightarrow \bv{v}^j}}$.

If $\bv{z}_{1:T}$ and $\bv{v}$ have been successfully disentangled, the difference between $\bv{z}_{1:T}^{\bv{v}^i \rightarrow \bv{v}^j}$ and $\bv{z}_{1:T}^i$ would be minimal because replacing the global factor should not affect the subsequently inferred local factor; and $\bv{v}^{\bv{v}^i \rightarrow \bv{v}^j}$ should be close to $\bv{v}^{j}$ in order to faithfully manifest the swapping.
Similarly, if we replace $\bv{z}_{1:T}$ instead, difference between $\bv{v}^{\bv{z}_{1:T}^i \rightarrow \bv{z}_{1:T}^j}$ and $\bv{v}^{i}$ is expected to be small; and $\bv{z}_{1:T}^{\bv{z}_{1:T}^i \rightarrow \bv{z}_{1:T}^j}$ should be close to $\bv{z}_{1:T}^j$.

We can impose the desired properties of factor invariance as well as the rendering of the target factors by introducing the following terms to \refeq{eq:dsae_elbo}:
\begin{align}
    & - \KLD{
    \qd{\bv{v} | \bv{x}^{\fs{\bv{v}}{i}{j}}_{1:T}}
    }{
    \qd{\bv{v} | \bv{x}_{1:T}^j}
    }, \label{swap_g_v} \\
    & - \KLD{
    \qd{\bv{z}_{1:T} | \bv{x}^{\fs{\bv{v}}{i}{j}}_{1:T}}
    }{
    \qd{\bv{z}_{1:T} | \bv{x}_{1:T}^i}
    }, \label{swap_g_z} \\
    & - \KLD{
    \qd{\bv{v} | \bv{x}^{\fs{\bv{z}_{1:T}}{i}{j}}_{1:T}}
    }{
    \qd{\bv{v} | \bv{x}_{1:T}^i}
    }, \label{swap_l_g} \text{and} \\
    & - \KLD{
    \qd{\bv{z}_{1:T} | \bv{x}^{\fs{\bv{z}_{1:T}}{i}{j}}_{1:T}}
    }{
    \qd{\bv{z}_{1:T} | \bv{x}_{1:T}^j}
    }. \label{swap_l_z}
\end{align}
By maximising these terms, we encourage invariance of the local and global latent variables through \refeq{swap_g_z} and \refeq{swap_l_g}, respectively.
Meanwhile, posteriors of the replaced factors are regularised to follow the target posteriors through \refeq{swap_g_v} and \refeq{swap_l_z}.

In practice, we pair each input sequence $i$ in a mini-batch with a randomly sampled input sequence $j$ from the same mini-batch, and perform the above-mentioned scheme of inference, replacement, decoding, and inference.
Note that we do not require any form of supervision or data-augmentation.
While the above terms encourage meaningful behaviour, they can still be minimised with a trivial global latent space, which is undesired.
Thus, the two-stage training plays a crucial role in obtaining robust disentanglement.
Further, note that the individual terms above vary in terms of magnitude and thus importance to the gradient and so could benefit from balancing. However, we found scaling them unnecessary for the success of disentanglement, and leave this study for future work.

To summarise, TS-DSAE constitutes a two-stage training framework that facilitates the exploitation of additional divergences to achieve robust unsupervised disentanglement, which we empirically verify in \secref{sec:result}.

\section{Related Work}\label{sec:related}
The assumption of a sequence being generated by a stationary global factor and a temporally changing local factor to achieve unsupervised disentanglement was used before.
FHVAE~\cite{hsu2017fhvae} constructs a hierarchical prior where each input is governed by a sequence-level prior on top of a segment-level prior.
Our two-stage training framework shares the spirit, with the main difference being that we leverage the strong bottleneck during the constrained training to naturally promote a global information-rich posterior which can be directly used as the sequence-level prior for the complete model training stage.
On the other hand, FHVAE initialises and learns the prior from scratch, which lacks a stronger inductive bias and a discriminative objective function is reported to be helpful.
Also, learning of the sequence-level priors is amortised by the global encoder in our model, whereby memory consumption does not scale with the number of training data as in FHVAE.

The vanilla DSAE~\cite{Li2018dsae} is proposed as an elegant minimalistic model to achieve disentanglement, as shown in \figref{fig:dsae}.
However, we demonstrate its tendency to collapse the global latent space in \secref{sec:result}, which is likely due to the over-simplified standard Gaussian prior.
R-WAE~\cite{Han2021RWAE} minimises the Wasserstein distance between the aggregated posterior and the prior instead, estimated by maximum mean discrepancy or generative adversarial networks, either of which is not trivial in terms of parameter tuning and optimisation.
S3-VAE~\cite{Zhu2020S3VAESS} and C-DSVAE~\cite{Bai2021cdsvae} exploit self-supervised learning and employ either domain-specific ad-hoc loss functions or data augmentation.
The proposed TS-DSAE is free from any form of supervision, adversarial training, or domain-dependent data augmentation.

VDSM adopts a pre-training stage as well as a scheme of KL-annealing to promote usage of the global latent space~\cite{Vowels2021VDSMUV}, which is similar to our constrained training.
The main differences, however, are that we train only the global variable during ``pre-trainig'', and avoid KL-annealing to save the tuning efforts.
Further, VDSM employs $n$ decoders, each of which is  responsible for a unique identity of a video object, where $n$ is set manually depending on the dataset.
This makes it less general, requires rather heavy computation, and might complicate the optimisation process.
Lezama~\shortcite{lezama2019jacobian} proposes a progressive autoencoder-based framework to tackle the ``shortcut problem'' for static data.
The framework first trains a network with a low capacity latent space in order to learn the factors of interest, and subsequently increases the latent space capacity to improve data reconstruction.
The final model utilises supervision from human annotations to learn the factors of interest.
Our two-stage training shares a similar idea, but differs in that TS-DSAE operates without any supervision and models sequential data.

Our constrained training stage is also reminiscent of multi-view representation learning.
For example, VCCA~\cite{Wang2016dvcca} formulates a model that samples different views of a common object 
from distributions conditioned on a shared latent variable.
NestedVAE~\cite{Vowels2020NestedVAEIC} learns the common factors using staged information bottlenecks by training a low-level VAE given the latent space derived from a high-level VAE.
In our model, given an input sequence, we treat multiple time frames as the different ``views'' of a common underlying factor which is the global factor.

There has been a lack of exploration in unsupervised disentangled representation for music audio.
Both Luo \textit{et al.}~\shortcite{Luo2020UnsupervisedDO} 
and C{\'i}fka \textit{et al.}~\shortcite{cifka2021vqvae} exploit self-supervised learning to decorrelate instrument pitch and timbre.
Similar to our work, the latter models monophonic melodies.
Yet, it employs pitch-shifting which is domain-dependent, and constrains the local capacity by learning discrete latent variables which might pose optimisation challenges.
We maintain the simplicity of DSAE and improve the robustness in a simple yet effective way, which is not limited to any certain modality.

\section{Experimental Setup}\label{sec:experiment}
\subsection{Datasets}
We consider both an artificial and a real-world music audio dataset.
The former facilitates the control over the underlying factors of variation, while the latter demonstrates applicability of the proposed model to realistic data.

\paragraph{dMelodies.}
The artificial dataset is compiled by synthesising audio from monophonic symbolic music gathered from dMelodies~\cite{pati2020dmel}.
Each melody is a two-bar sequence with 16 eighth notes, subject to several global factors, i.e., tonic, scale, and octave, and local factors, i.e., direction of arpeggiation, and rhythm.
In order to facilitate analysis, we normalise the global factors by considering only the melodies of C Major in the fourth octave.
We also discard melodies starting or ending with the rest note to avoid spurious amplitude values and boundaries during audio synthesis with FluidSynth.\footnote{https://www.fluidsynth.org/}
We randomly pick 3k samples from the remaining melodies which are then split into 80\% training and 20\% validation sets, and synthesise audio of sampling rate 16kHz using sound fonts of violin and trumpet from MuseScore\_General.sf3.\footnote{https://musescore.org/en/handbook}
The amplitude of each audio sample is normalised with respect to its maximum value.
The number of samples rendered with the two instruments is uniformly distributed.

\paragraph{URMP.}
For the real-world audio recordings, we select the violin and trumpet tracks from the URMP dataset~\cite{urmp}.
We follow the preprocessing by Hayes \textit{et al.}~\shortcite{Hayes2021nwe}, where the amplitude of each audio recording, resampled to 16kHz, is normalised in a corpus-wide fashion for each instrument subset.
The audio samples are then divided into four-second segments, and segments with mean pitch confidence lower than 0.85 are discarded, as assessed by the full CREPE model~\cite{Kim2018CrepeAC}, a state-of-the-art pitch extractor.
The process results in 1,545 violin and 534 trumpet samples in the training set, and 193 violin and 67 trumpet samples for validation.

Note that for both datasets, we expect the underlying local and global factors to be melody and instrument identity, respectively.
We transform the audio samples and represent the data as log-amplitude mel-spectrogram with 80 mel filter banks, derived from a short-time Fourier transform with a 128ms Hann window and 16ms hop, leading to $\bv{x}_{1:T} \in \mathbb{R}^{80 \times 251}$.

\subsection{Implementation}
\paragraph{Architecture.}
We study the two models proposed in the original DSAE~\cite{Li2018dsae}, ``factorised $q$'' and ``full $q$'' as shown in \figref{fig:dsae}.
We use \texttt{net-[layers]} to denote architectures of modules, where \texttt{net} indicates types of the network, and \texttt{[layers]} is a list specifying the numbers of neurons at each layer. 
\texttt{Tanh} is used as the non-linear activation between layers of FCNs, and we use long short-term memory (LSTM) for RNNs.
If a Gaussian parameterisation layer follows, we append the notation \texttt{Gau-L} which encompasses two linear layers with parameters $\bv{w}_1$ and $\bv{w}_2$ projecting the output hidden states $\bv{h}$ to $\mu_{\bv{w}_1}(\bv{h}) \in \mathbb{R}^L$ and $\log\sigma_{\bv{w}_2}^2(\bv{h}) \in \mathbb{R}^L$, respectively, where the Gaussian variable living in an $L$-dimensional space is then sampled from $\gauss{\mu_{\bv{w}_1}(\bv{h})}{\mathrm{diag}(\sigma_{\bv{w}_2}^2(\bv{h}))}$.

For factorised $q$, we implement the global encoder $\qd{\bv{v} | \bv{x}_{1:T}}$ as \texttt{FCN-[64,64]-Avg-Gau-{16}}, where \texttt{Avg} denotes average pooling across the time-axis, and we keep the size of $\bv{v}$ fixed as 16 across our main experiments; and the local encoder $\qd{\bv{z}_{1:T} | \bv{x}_{1:T}}$ as \texttt{FCN-[64,64]-Gau-\{8,16,32\}}, where we investigate different sizes of $\bv{z}_{1:T}$.
For the transition network $\pd{\bv{z}_t | \bv{z}_{<t}}$, we use \texttt{RNN-[32,32]-Gau-\{8,16,32\}}.
The decoder $\pd{\bv{x}_{t} | \bv{z}_t, \bv{v}}$ is \texttt{FCN-[64,64]-Gau-80} taking as input the concatenation of $\bv{z}_{1:T}$ and time-axis broadcast $\bv{v}$.
Note that, following the convention of VAEs, the Gaussian layer of the decoder parameterises $\gauss{\mu_{\bv{w}_1}(\bv{h})}{\bv{1}}$ which evaluates the likelihood $\pd{\bv{x}_{t} | \bv{z}_t, \bv{v}}$ as the squared L2-norm between the output of the decoder and $\bv{x}_{t}$.

For full $q$, $\qd{\bv{v} | \bv{x}_{1:T}}$ follows that of factorised $q$; and $\qd{\bv{z}_{1:T} | \bv{x}_{1:T}, \bv{v}}$ corresponds to \texttt{biRNN-[64,64]-Gau-\{8,16,32\}} which takes as input the concatenation of $\bv{x}_{1:T}$ and time-axis broadcast $\bv{v}$ inferred from $\qd{\bv{v} | \bv{x}_{1:T}}$.
\texttt{biRNN} denotes a bi-LSTM, where the outputs of the forward and backward LSTM are averaged along the time-axis before the Gaussian layer.
Both the transition network and decoder follow those of factorised $q$.

\paragraph{Optimisation.}
Our implementation is based on \texttt{PyTorch} \texttt{v1.9.0}
and we use ADAM~\cite{Kingma2015AdamAM} with default parameters $lr = 0.001$, and $[\beta_1, \beta_2]=[0.9, 0.999]$ without weight decay.
We use a batch size of 128, and train the models for 4k epochs at most; we employ early stopping if \refeq{eq:dsae_elbo} obtained from the validation set stops improving for 300 epochs.
For the models adopting the proposed two-stage training frameworks presented in~\secref{sec:method}, we set the number of epochs for the first stage $C=300$ for all cases, to which we find the performance insensitive.

\section{Experiments and Results}\label{sec:result}
We consider three baseline methods: 1) \textit{DSAE}; 2) \textit{DSAE-f}, where we employ the constrained training and freeze the global encoder after $C$ epochs; and 3) \textit{TS-DSAE w/o regs}, where we adopt the two-stage training framework without introducing the four terms from \secref{sec:klds}.
We do not include the models mentioned in \secref{sec:related} \cite{Zhu2020S3VAESS,Bai2021cdsvae,Han2021RWAE} which is left for future work, because the main focus is to improve upon DSAE with minimum modifications, and thus provide a superior backbone model which can be complementary with the existing methods.

 \begin{figure}[!t]
\centering  
 \includegraphics[width=0.95\columnwidth]{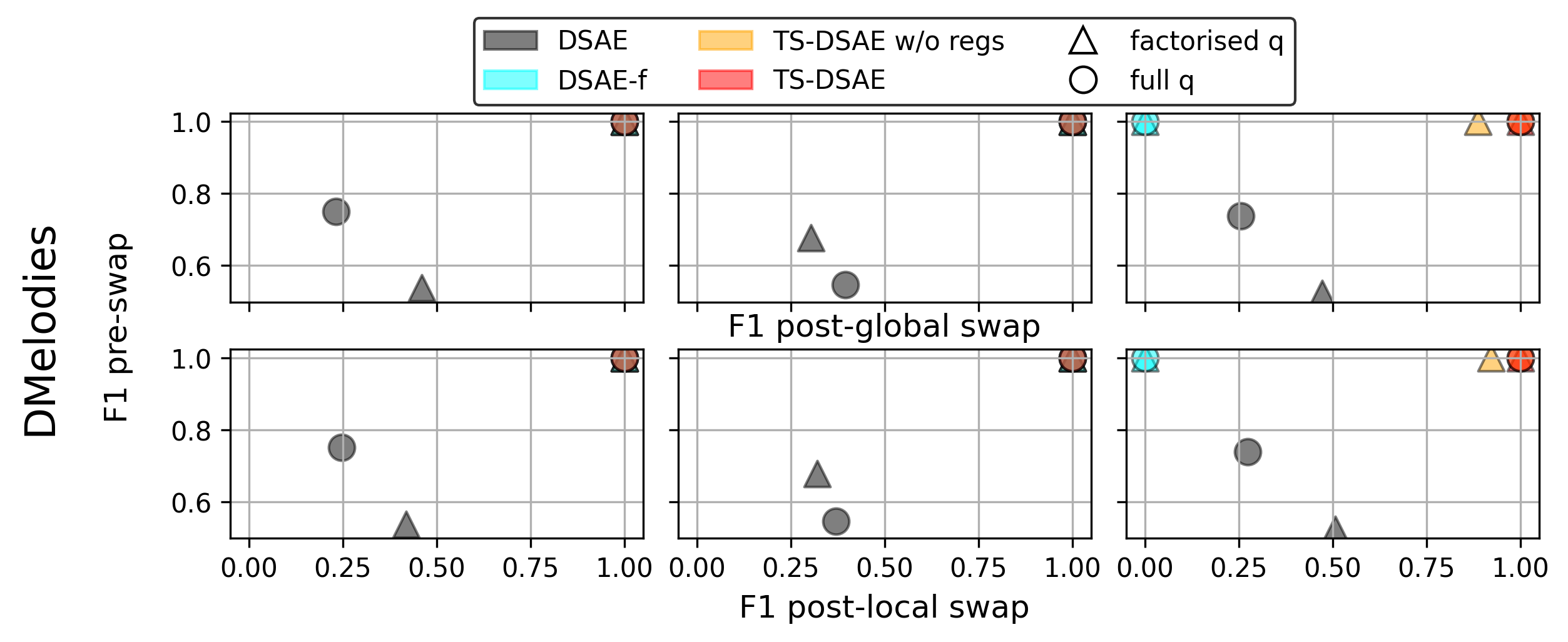}
 \bigbreak
 \includegraphics[width=0.95\columnwidth]{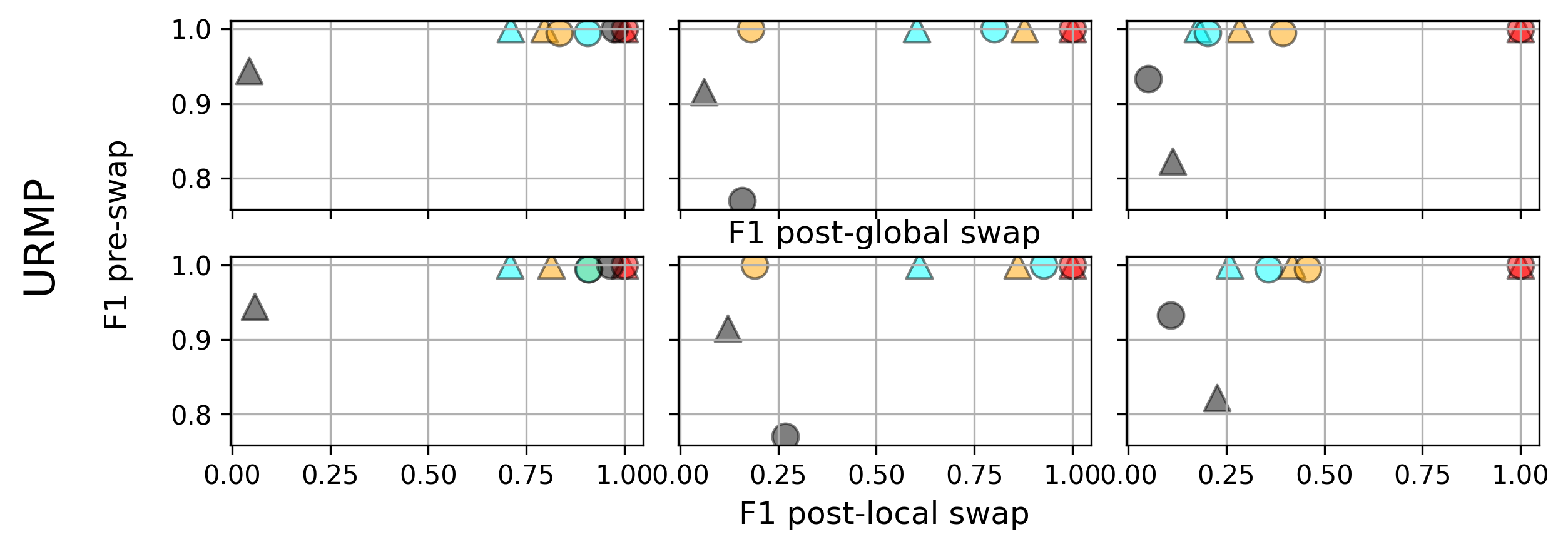}
\vspace{-5pt}
 \caption{Macro F1 score of instrument classification derived from applying LDA to the global latent space.
 Size of the local latent space increases from left to right columns, 8, 16, and 32, respectively.
 See \secref{sec:LDA} for details.
 }
 \label{fig:main_result_global}
 \vspace{-10pt}
\end{figure}

\subsection{Instrument Classification}\label{sec:LDA}
We first evaluate disentanglement through the lens of instrument classification.
In particular, we train a linear discriminant analysis (LDA) classifier taking as inputs $\bv{v} ~\sim \qd{\bv{v} | \bv{x}_{1:T}}$, the global latent variables sampled from a learnt model, derived from the training set, and evaluate its classification accuracy for instrument identity in terms of the macro F1-score on the validation set.
We pair each sequence $i$ from the validation set with another sequence $j$ recorded with the other instrument, and perform the scheme of inference, replacement, decoding, and inference.
Following the notation in \secref{sec:klds}, $\bv{v}^{\fs{\bv{v}}{i}{j}}$ should be predictive of the instrument of sample $j$; while $\bv{v}^{\fs{\bv{z}_{1:T}}{i}{j}}$ should reflect the original instrument of sample $i$.
We report three metrics including accuracy before the replacement (pre-swap), after replacing $\bv{v}$ (post-global swap), and after replacing $\bv{z}_{1:T}$ (post-local swap). Note that we use the mean parameters of the Gaussian posterior $\qd{\bv{v} | \bv{x}_{1:T}}$ to train the LDA.

The results are summarised in \figref{fig:main_result_global}.
The proposed TS-DSAE (red), with either factorised or full $q$, is consistently located at the top right corner of the plot, across all the sizes of the local latent space.
This indicates its robust disentanglement as well as a linearly separable global latent space.
From the left to right column, the competing methods DSAE-f (cyan) and TS-DSAE without the additional regularisations (orange) move from top right to left-hand side of the plot, showing the inclination for a collapsed global latent space with the increased local latent capacity.
Being located at lower left of the plot, DSAE (gray) attains the worst performance in most configurations.
This highlights the issue of positing the standard Gaussian prior in the global latent space.

The overall high pre-swap and low post-swap F1 especially towards high-dimensional $\bv{z}_{t}$ implies that the decoder tends to ignore $\bv{v}$, even though the mean parameter of $\qd{\bv{v} | \bv{x}_{1:T}}$ is discriminative w.r.t. the instrument identity.
The competing models appear to suffer the most from the size of $\bv{z}_{t}$ as a large local latent space can easily capture all the necessary information for reconstruction.

\begin{figure}[!t]
\centering  
 \includegraphics[width=0.75\columnwidth]{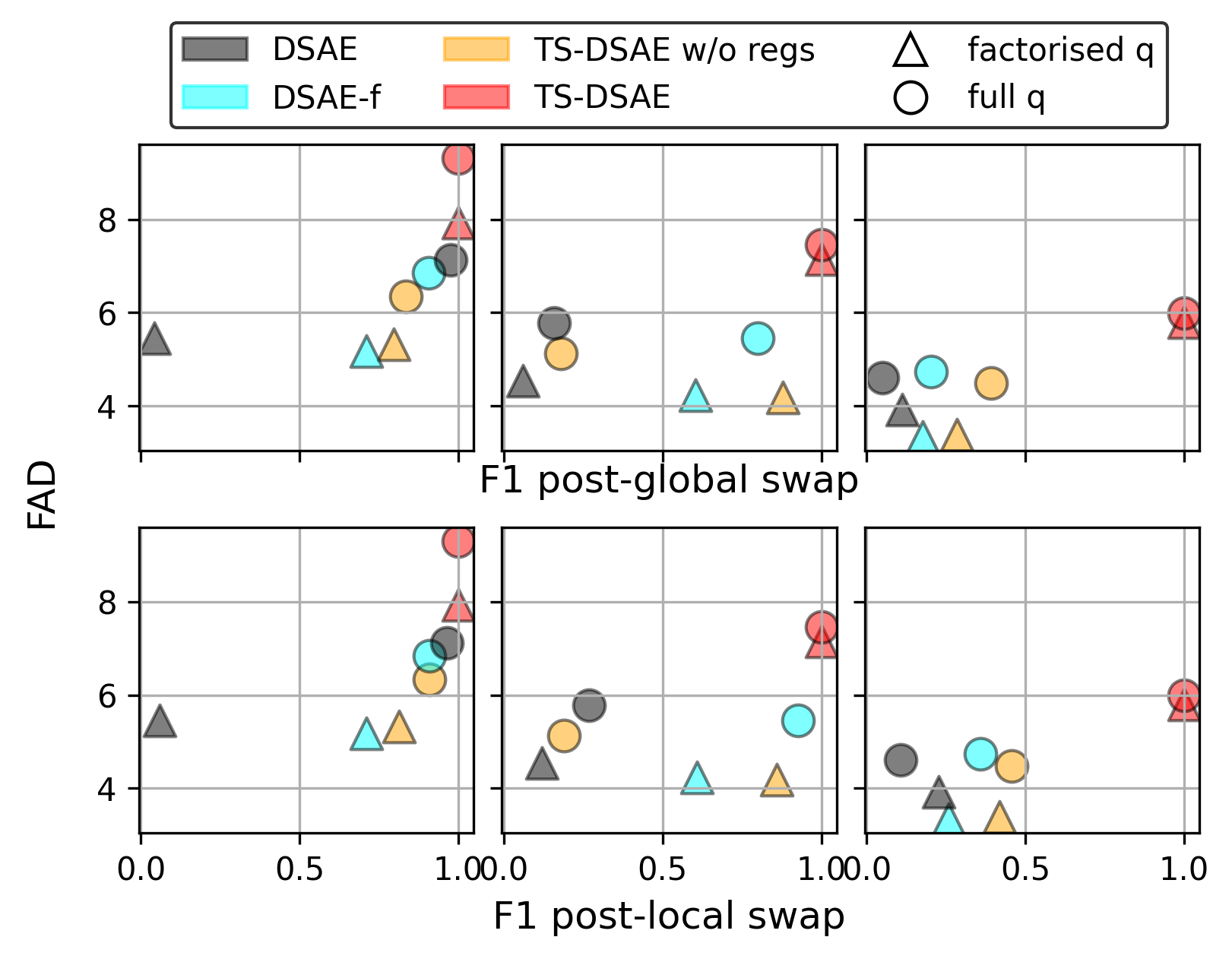}
  \vspace{-5pt}
 \caption{FAD (the lower the better) of reconstruction versus macro F1 score for instrument classification, evaluated using URMP.
 See \secref{sec:FAD} for details.
 }
 \label{fig:main_result_fad}
 \vspace{-10pt}
\end{figure}

\subsection{Reconstruction Quality}\label{sec:FAD}
We examine the trade-off between disentanglement and reconstruction in terms of Fréchet Audio Distance (FAD)~\cite{Kilgour2019fad} which is reported to correlate with auditory perception.
We only report the results for URMP in \figref{fig:main_result_fad} as both datasets reach a similar summary.
As expected, FAD is improved with increasing $\bv{z}_{t}$ dimension.
However, TS-DSAE is the only model that overcomes the trade-off, in the sense that competing models lose their ability to disentangle (move from right to left of the plot) with the improved FAD.



\subsection{Raw Pitch Accuracy}

In this section, we evaluate $\bv{z}_{1:T}$ by applying the full CREPE model~\cite{Kim2018CrepeAC} to audio re-synthesised from the mel-spectrogram.
The conversion is done by \texttt{InverseMelScale} and \texttt{GriffinLim} accessible from \texttt{torchaudio} \texttt{v0.9.0}.
Using the notation from \secref{sec:klds}, we extract pitch contours from reconstructed samples (pre-swap), $\bv{x}_{1:T}^{\bv{v}^i \rightarrow \bv{v}^j}$ (post-global swap) which is supposed to mirror the pitch contour of $\bv{x}_{1:T}^i$, and $\bv{x}_{1:T}^{\bv{z}_{1:T}^i \rightarrow \bv{z}_{1:T}^j}$ (post-local swap) which is supposed to follow the pitch contour of $\bv{x}_{1:T}^j$.
Note that for models with trivial $\bv{v}$, the accuracy of post-global swap will remain high as the decoder is independent of $\bv{v}$.
We extract pitch contours from the input data as the ground-truth and report the raw pitch accuracy (RPA) with a 50-cent threshold~\cite{Salamon2014MelodyEF}.

\begin{figure}[!t]
\centering  
 \includegraphics[width=0.75\columnwidth]{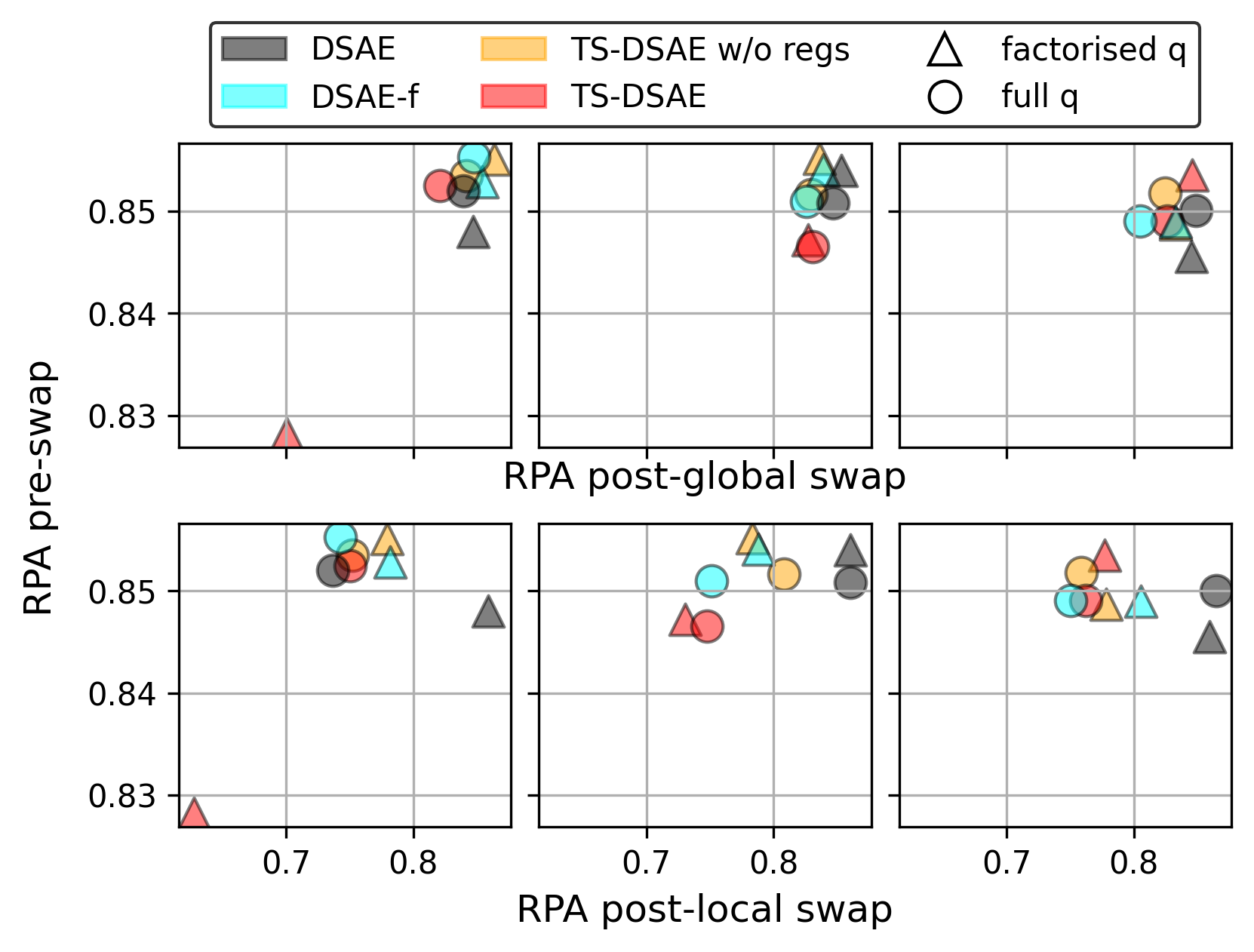}
 \vspace{-5pt}
 \caption{
 RPA assessed using CREPE on URMP.
 }
 \label{fig:main_result_local}
 \vspace{-10pt}
\end{figure}

We report the results with URMP in \figref{fig:main_result_local}. 
TS-DSAE consistently improves with the increasing size of $\bv{z}_t$ in terms of RPA.
Except for post-local swap, TS-DSAE performs comparably with the competing models towards the larger $\bv{z}_t$, and achieves disentanglement at once.

\subsection{Richer Decoders}\label{sec:rich_decoder}
To mitigate the trade-off, we further construct and evaluate a richer decoder where the reconstruction of $\bv{x}_t$ is conditioned on $\bv{z}_{1:T}$, i.e., the entire sequence of local latent variables, instead of $\bv{z}_{t}$.
We set the size of $\bv{z}_{t}$ to 16, and the inference network to factorised $q$, and compare DSAE, TS-DSAE, and the TS-DSAE augmented with the enriched decoder.

\begin{figure}[!b]
\centering  

 \includegraphics[width=\columnwidth]{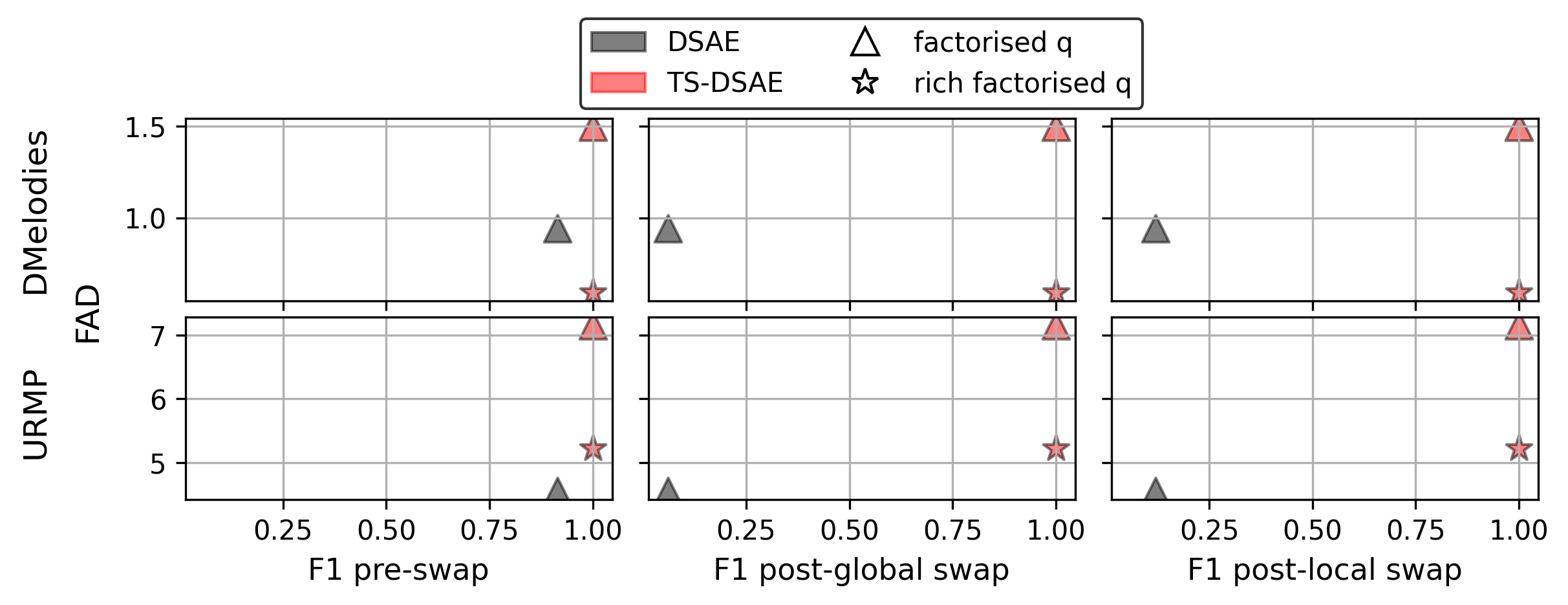}

\vspace{-5pt}
\caption{FAD (the lower the better) against disentanglement in terms of instrument classification.}
 \label{fig:rich}
\end{figure}

As shown in \figref{fig:rich}, the enriched model maintains the perfect accuracy for instrument classification for both datasets, while improving FAD over its counterpart with the factorised decoder.
Note that using dMelodies, the model outperforms DSAE equipped with the factorised decoder in terms of FAD.

We leave the evaluation for the full range of configurations for future work, including autoregressive decoders that could cause posterior collapse even for vanilla VAEs.

 \begin{figure}[!t]
\centering  
 \includegraphics[width=0.9\columnwidth]{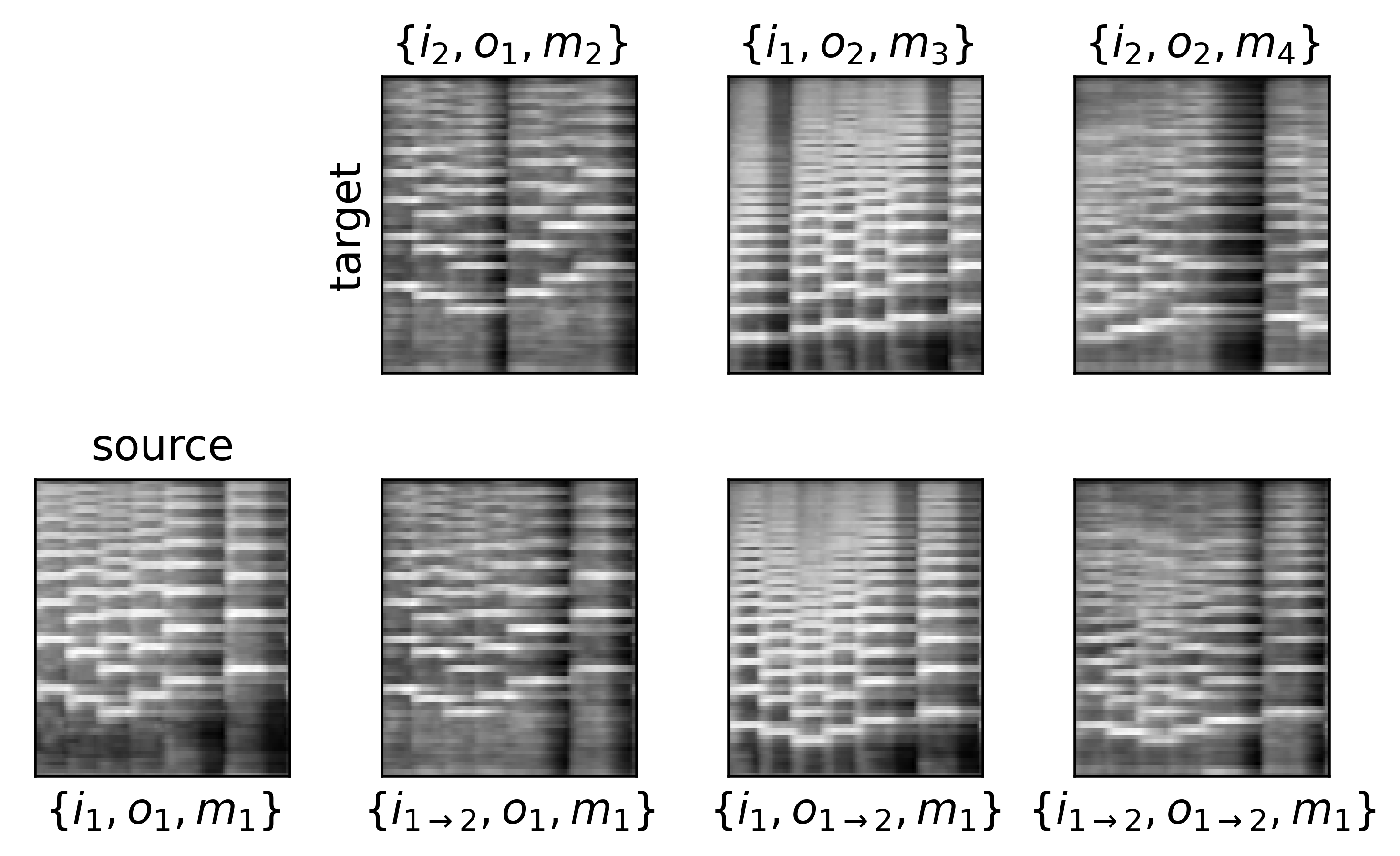}
 \vspace{-5pt}
\caption{Global latent replacement using the top three samples as the targets and the sample at the bottom left as the source.}
 \label{fig:multiple}
 \vspace{-10pt}
\end{figure}

\subsection{Multiple Global Factors}

We now consider both the fourth and fifth octaves when synthesising the dMelodies dataset, introducing octave number as the other global factor of variation in addition to instrument identity.
We train the decoder-enriched TS-DSAE described in \secref{sec:rich_decoder}, and show the results in \figref{fig:multiple}.
In particular, we replace $\bv{v}$ inferred from the source at the lower left, with that derived from one of the three targets displayed in the top row, and generate novel samples shown from the second to last columns of the bottom row.

We use $\{i, o, m\}$ to denote the instrument, octave, and melody of each sample, respectively.
For example, the source $\{i_1, o_1, m_1\}$ and the first target $\{i_2, o_1, m_2\}$ share the same octave but differ in the instrument, characterised by the spectral distribution along the frequency axis.
As a result of replacing $\bv{v}$, the target instrument $i_2$ is manifested in the outcome $\{i_{1\rightarrow2}, o_1, m_1\}$, while the octave remains unchanged.
Similarly, the second target $\{i_1, o_2, m_3\}$ differs from the source with the octave, characterised by the level of pitch contour; therefore, swapping $\bv{v}$ only transforms the octave for the output $\{i_1, o_{1\rightarrow2}, m_1\}$.
Finally, the sample $\{i_{1\rightarrow2}, o_{1\rightarrow2}, m_1\}$ results from using the target $\{i_2, o_2, m_4\}$ that does not share any of the attributes with the source, where both the instrument and octave are converted.
Importantly, the source melody $m_1$ remains intact in the three transformed samples, suggesting the global-local disentanglement.

\section{Conclusion and Future Work}\label{sec:conclusion}
We have proposed TS-DSAE, a robust framework for unsupervised sequential data disentanglement, which has been shown to consistently work over a wide range of settings.
\footnote{The implementation and audio samples are accessible from https://github.com/yjlolo/dSEQ-VAE.}
Our evaluation focuses on the ability to robustly achieve disentanglement, and we leave evaluations on multi-modal data generation from unconditional prior sampling for future work.
We would also like to verify the applicability of TS-DSAE to modalities beyond the music audio datasets. 

Despite the drastic increase in robustness, the difficulty of balancing disentanglement and reconstruction remains challenging~\cite{lezama2019jacobian}.
Scaling the regularisation terms differently might be helpful as mentioned in ~\secref{sec:klds}.
Moreover, DSAEs probabilistic graphical model forces the input sequence to have a single global latent variable fixed over time, which could be too restrictive for more general use cases where sequences do not have stationary factors but ones that evolve slowly. 
Therefore, adopting a hierarchy of latent variables encoding information at 
multiple frame rates~\cite{Saxena2021ClockworkVA} can be a favorable relaxation of DSAEs.
A potential extension of our two-stage training is to have multiple stages of constrained training with progressively larger network capacity, thereby accommodating the said hierarchy, which can also be seen as a temporal extension of Li \textit{et al.}~\shortcite{Li2020ProgressiveLA}.

\section*{Acknowledgments}

The first author is a research student at the UKRI CDT in AI and Music, supported by Spotify.

\bibliographystyle{named}
\bibliography{ijcai22}

\begin{thebibliography}{}

\bibitem[\protect\citeauthoryear{Bai \bgroup \em et al.\egroup
  }{2021}]{Bai2021cdsvae}
Junwen Bai, Weiran Wang, and Carla Gomes.
\newblock Contrastively disentangled sequential variational autoencoder.
\newblock {\em Advances in Neural Information Processing Systems}, 2021.

\bibitem[\protect\citeauthoryear{Bengio}{2013}]{bengio2013deep}
Yoshua Bengio.
\newblock Deep learning of representations: Looking forward.
\newblock In {\em Proceedings of the International Conference on Statistical
  Language and Speech Processing}, 2013.

\bibitem[\protect\citeauthoryear{C{\'i}fka \bgroup \em et al.\egroup
  }{2021}]{cifka2021vqvae}
Ondřej C{\'i}fka, Alexey Ozerov, Umut Şimşekli, and Gaël Richard.
\newblock Self-supervised {VQ-VAE} for one-shot music style transfer.
\newblock In {\em Proceedings of the International Conference on Acoustics,
  Speech and Signal Processing}, 2021.

\bibitem[\protect\citeauthoryear{Han \bgroup \em et al.\egroup
  }{2021}]{Han2021RWAE}
Jun Han, Martin~Renqiang Min, Ligong Han, Li~Erran Li, and Xuan Zhang.
\newblock Disentangled recurrent {W}asserstein autoencoder.
\newblock In {\em Proceedings of the International Conference on Learning
  Representations}, 2021.

\bibitem[\protect\citeauthoryear{Hayes \bgroup \em et al.\egroup
  }{2021}]{Hayes2021nwe}
Ben Hayes, Charalampos Saitis, and György Fazekas.
\newblock Neural waveshaping synthesis.
\newblock In {\em Proceedings of the International Society for Music
  Information Retrieval Conference}, 2021.

\bibitem[\protect\citeauthoryear{Hsu \bgroup \em et al.\egroup
  }{2017}]{hsu2017fhvae}
Wei-Ning Hsu, Yu~Zhang, and James Glass.
\newblock Unsupervised learning of disentangled and interpretable
  representations from sequential data.
\newblock {\em Advances in Neural Information Processing Systems}, 2017.

\bibitem[\protect\citeauthoryear{Khurana \bgroup \em et al.\egroup
  }{2019}]{Khurana2019fdmm}
Sameer Khurana, Shafiq~Rayhan Joty, Ahmed Ali, and James Glass.
\newblock A factorial deep {M}arkov model for unsupervised disentangled
  representation learning from speech.
\newblock In {\em Proceedings of the International Conference on Acoustics,
  Speech and Signal Processing}, 2019.

\bibitem[\protect\citeauthoryear{Kilgour \bgroup \em et al.\egroup
  }{2019}]{Kilgour2019fad}
Kevin Kilgour, Mauricio Zuluaga, Dominik Roblek, and Matthew Sharifi.
\newblock Fréchet audio distance: A reference-free metric for evaluating music
  enhancement algorithms.
\newblock In {\em Proceedings of INTERSPEECH}, 2019.

\bibitem[\protect\citeauthoryear{Kim \bgroup \em et al.\egroup
  }{2018}]{Kim2018CrepeAC}
Jong~Wook Kim, Justin Salamon, Peter~Qi Li, and Juan~Pablo Bello.
\newblock {CREPE}: A convolutional representation for pitch estimation.
\newblock In {\em Proceedings of the International Conference on Acoustics,
  Speech and Signal Processing}, 2018.

\bibitem[\protect\citeauthoryear{Kingma and Ba}{2014}]{Kingma2015AdamAM}
Diederik~P. Kingma and Jimmy Ba.
\newblock Adam: A method for stochastic optimization.
\newblock {\em arXiv preprint arXiv:1412.6980}, 2014.

\bibitem[\protect\citeauthoryear{Kingma and Welling}{2014}]{kingma2013auto}
Diederik~P Kingma and Max Welling.
\newblock Auto-encoding variational bayes.
\newblock In {\em Proceedings of the International Conference on Learning
  Representations}, 2014.

\bibitem[\protect\citeauthoryear{Lezama}{2019}]{lezama2019jacobian}
José Lezama.
\newblock Overcoming the disentanglement vs reconstruction trade-off via
  {Jacobian} supervision.
\newblock In {\em Proceedings of the International Conference on Learning
  Representations}, 2019.

\bibitem[\protect\citeauthoryear{Li and Mandt}{2018}]{Li2018dsae}
Yingzhen Li and Stephan Mandt.
\newblock Disentangled sequential autoencoder.
\newblock In {\em Proceedings of the International Conference on Machine
  Learning}, 2018.

\bibitem[\protect\citeauthoryear{Li \bgroup \em et al.\egroup }{2019}]{urmp}
Bochen Li, Xinzhao Liu, Karthik Dinesh, Zhiyao Duan, and Gaurav Sharma.
\newblock Creating a multitrack classical music performance dataset for
  multimodal music analysis: Challenges, insights, and applications.
\newblock {\em IEEE Transactions on Multimedia}, 2019.

\bibitem[\protect\citeauthoryear{Li \bgroup \em et al.\egroup
  }{2020}]{Li2020ProgressiveLA}
Zhiyuan Li, Jaideep~Vitthal Murkute, Prashnna~Kumar Gyawali, and Linwei Wang.
\newblock Progressive learning and disentanglement of hierarchical
  representations.
\newblock In {\em Proceedings of the International Conference on Learning
  Representations}, 2020.

\bibitem[\protect\citeauthoryear{Locatello \bgroup \em et al.\egroup
  }{2019}]{locatello2018challenging}
Francesco Locatello, Stefan Bauer, Mario Lucic, Gunnar R{\"a}tsch, Sylvain
  Gelly, Bernhard Sch{\"o}lkopf, and Olivier Bachem.
\newblock Challenging common assumptions in the unsupervised learning of
  disentangled representations.
\newblock In {\em Proceedings of the International Conference on Machine
  Learning}, 2019.

\bibitem[\protect\citeauthoryear{Luo \bgroup \em et al.\egroup
  }{2020}]{Luo2020UnsupervisedDO}
Yin-Jyun Luo, Kin~Wai Cheuk, Tomoyasu Nakano, Masataka Goto, and Dorien
  Herremans.
\newblock Unsupervised disentanglement of pitch and timbre for isolated musical
  instrument sounds.
\newblock In {\em Proceedings of the International Society for Music
  Information Retrieval Conference}, 2020.

\bibitem[\protect\citeauthoryear{Pati \bgroup \em et al.\egroup
  }{2020}]{pati2020dmel}
Ashis Pati, Siddharth Gururani, and Alexander Lerch.
\newblock {dMelodies}: {A} music dataset for disentanglement learning.
\newblock In {\em Proceedings of the International Society for Music
  Information Retrieval Conference}, 2020.

\bibitem[\protect\citeauthoryear{Salamon \bgroup \em et al.\egroup
  }{2014}]{Salamon2014MelodyEF}
Justin Salamon, Emilia G{\'o}mez, Daniel~P.~W. Ellis, and Ga{\"e}l Richard.
\newblock Melody extraction from polyphonic music signals: Approaches,
  applications, and challenges.
\newblock {\em IEEE Signal Processing Magazine}, 2014.

\bibitem[\protect\citeauthoryear{Saxena \bgroup \em et al.\egroup
  }{2021}]{Saxena2021ClockworkVA}
Vaibhav Saxena, Jimmy Ba, and Danijar Hafner.
\newblock Clockwork variational autoencoders.
\newblock {\em Advances in Neural Information Processing Systems}, 2021.

\bibitem[\protect\citeauthoryear{Vowels \bgroup \em et al.\egroup
  }{2020}]{Vowels2020NestedVAEIC}
Matthew~James Vowels, Necati~Cihan Camg{\"o}z, and Richard Bowden.
\newblock {NestedVAE}: Isolating common factors via weak supervision.
\newblock In {\em Proceedings of the International Conference on Computer
  Vision and Pattern Recognition}, 2020.

\bibitem[\protect\citeauthoryear{Vowels \bgroup \em et al.\egroup
  }{2021}]{Vowels2021VDSMUV}
Matthew~James Vowels, Necati~Cihan Camgoz, and Richard Bowden.
\newblock {VDSM}: Unsupervised video disentanglement with state-space modeling
  and deep mixtures of experts.
\newblock In {\em Proceedings of the International Conference on Computer
  Vision and Pattern Recognition}, 2021.

\bibitem[\protect\citeauthoryear{Wang \bgroup \em et al.\egroup
  }{2016}]{Wang2016dvcca}
Weiran Wang, Honglak Lee, and Karen Livescu.
\newblock Deep variational canonical correlation analysis.
\newblock {\em arXiv preprint arxiv:1610.03454}, 2016.

\bibitem[\protect\citeauthoryear{Zhu \bgroup \em et al.\egroup
  }{2020}]{Zhu2020S3VAESS}
Yizhe Zhu, Martin~Renqiang Min, Asim Kadav, and Hans~Peter Graf.
\newblock {S3VAE}: Self-supervised sequential {VAE} for representation
  disentanglement and data generation.
\newblock In {\em Proceedings of the International Conference on Computer
  Vision and Pattern Recognition}, 2020.

\end{thebibliography}
\end{document}